# Towards Scalable FMI-based Co-simulation of Wind Energy Systems Using PowerFactory

Arjen A. van der Meer [*], Rishabh Bhandia [*], Edmund Widl[†], Kai Heussen[‡], Cornelius Steinbrink[§],
Przemyslaw Chodura[¶], Thomas I. Strasser[†], Peter Palensky [*]

[*] Delft University of Technology, Delft, The Netherlands, {a.a.vandermeer, r.bhandia, p.palensky}@tudelft.nl
[†] AIT Austrian Institute of Technology, Vienna, Austria {edmund.widl, thomas.strasser}@ait.ac.at
[‡] Technical University of Denmark, Lyngby, Denmark kh@elektro.dtu.dk
[§] OFFIS – Institute for Information Technology, Oldenburg, Germany cornelius.steinbrink@offis.de
[¶] DNV GL, Arnhem, The Netherlands maui@mauiconsulting.co

*Abstract*—Due to the increased deployment of renewable energy sources and intelligent components the electric power system will exhibit a large degree of heterogeneity, which requires inclusive and multi-disciplinary system assessment. The concept of co-simulation is a very attractive option to achieve this; each domain-specific subsystem can be addressed via its own specialized simulation tool. The applicability, however, depends on aspects like standardised interfaces, automated case creation, initialisation, and the scalability of the co-simulation itself. This work deals with the inclusion of the Functional Mock-up Interface for co-simulation into the DIgSILENT PowerFactory simulator, and tests its accuracy, implementation, and scalability for the grid connection study of a wind power plant. The coupling between the RMS mode of PowerFactory and MATLAB/Simulink in a standardised manner is shown. This approach allows a straightforward inclusion of black-boxed modelling, is easily scalable in size, quantity, and component type.

## I. Introduction

Recent developments such as the policy around the energy transition, technological advancements, and the continual electrification of the energy system have lead to an unprecedented change of the corresponding power grids [1]. While not so long ago the power system comprised mainly fossil fuelled generation by rotating machinery, the present power system contains a considerable proportion of converter-interfaced components, such as wind power plants (WPPs). Simultaneously the overall system became more active; more components are controlled and able to communicate at much higher rates than with Supervisory Control and Data Acquisition (SCADA) systems.

The shift from a predominantly physical system to one increasingly dominated by controlled dynamics of inverters implies a change in the behaviour exhibited by the individual components as well as the system as a whole. The consequences of this massive grid integration of renewables and further roll-out of smart grid methods must be assessed by simulation studies, advanced hardware-in-the-loop (HIL) approaches, and lab-based testing. The European ERIGrid project aims to develop such assessment methods to holistically test and validate component and sub-systems, and therefore support the roll-out of new technologies and solutions [2].

For simulation assessment methods—the standard approach to test system dynamics in the transient stability time-frame of interest—it is common to simplify the concerned subsystems to such extent that the phenomena of interest to grid code compliance are still included into the overall model. For the power system this commonly yielded a fundamental frequency projection of the grid strength and inertial response through a Thévenin equivalent, whereas a WPP is represented by a generic dynamic model of a wind turbine generator (WTG), scaled up to the WPP rating.

This approach is not suitable for future electric power systems, which contain device and control interactions over a wide time-frame of interest. Simulating the entire power system, in which each individual WTG is considered by a separate electromagnetic transient (EMT) model, is not an option either because of computation and data constraints. How can this simulation challenge be resolved? Co-simulation is a very promising option here [3]; it allows to split the overall system into subsystems, where each subsystem is addressed by a specialized tool. A master algorithm then orchestrates the overall simulation (time stepping, data exchange, signal transformations, initialisation). Standards like the Functional Mock-up Interface (FMI) [4] allow a generic low-level interface between subsystems and enable black-boxed integration of vendor-specific turbine models.

This work shows how grid integration aspects of WPPs can be studied by co-simulating power system models developed in DIgSILENT PowerFactory with a set of WTG component and control models developed in MATLAB/Simulink. A key challenge in the division of simulation models at the grid level is that for power flow calculations, initialisation, and dynamic simulation a mutual (differential-algebraic) coupling is present, which needs to be handled automatically and correctly by the master orchestrator [5]. The static generator model of PowerFactory as the interface model between the grid and WTG models is applied. During runtime the algebraic coupling is resolved by employing a time-shifting between simulation time steps and executing the sub-models sequentially. During initialisation the coupling is handled by letting the WTG models dictate the active power infeed into the grid model.

The paper starts with introducing the implementation of the FMI for PowerFactory using the FMI++ toolbox, and its integration into the co-simulation orchestrator MOSAIK. A detailed explanation of the capabilities of FMI++ and MOSAIK in this context have been already discussed in [6]; this work will mainly focus on the implementation aspects. Second, the inclusion of the complex system behaviour into a formal test description as developed in the ERIGrid context will be discussed, thereby focusing on the park-level controls and fault-ride through implementation of the Type 4 generic WTG. The paper continues with the validity of the co-simulation approach by comparing the response of a monolithic PowerFactory simulation to a small-scale co-simulation, both considering the onshore WPP as an aggregated WTG model. Finally the scalability of the co-simulation setup is assessed by disaggregating the wind park into 32 individual WTGs while the cable array is modelled in detail in PowerFactory.

## II. FMI-BASED Co-SIMULATION FOR POWERFACTORY

### A. An FMI-compliant interface for PowerFactory

Until now, PowerFactory does not naturally provide any FMI-compliant functionality for model exchange or co-simulation. Instead, it comes with an interface that enables basic interactions with a simulation model [7], e.g., setting/retrieving values of parameters and variables, triggering power flow calculations and starting/stopping time-domain simulations. Furthermore, so-called *events* can be issued during runtime, which can change the system state at a specified point in simulation time. This mechanism can be applied to alter loads or the status of switches, providing the means for dynamic interaction at run-time required for co-simulation.

This functionality has been mapped to an FMI-compliant interface[1], which supports two types of simulations. In *quasi-static steady-state simulations* a power system's evolution with respect to time is captured by a series of power flow snapshots. *RMS simulations* allow to calculate the time-dependent dynamics of electro-mechanical models, including control devices.

This paper considers WTG dynamics and hence utilizes the RMS functionality of PowerFactory. In this case, interfacing PowerFactory relies on the *DIgSILENT Simulation Language* (DSL), which enables programming controllers and other common power system component models. PowerFactory allows to associate an object with a user-defined DSL model via a *composite model*. By sending a *parameter event* to such a DSL model, the model's input parameters can be changed. This change of input parameters can be easily propagated to the parameters of any object by connecting the DSL model with the corresponding object in a composite model. For sending events to a composite model during a co-simulation, a dedicated DSL block called *FMIAdapter* is provided. To use this DSL block, it must be included into a composite model, an example of which is shown in Fig. 1. To which blocks these events are sent is defined in the FMU export configuration.

[1] The FMI++ PowerFactory FMU Export Utility, available at: http://powerfactory-fmu.sourceforge.net

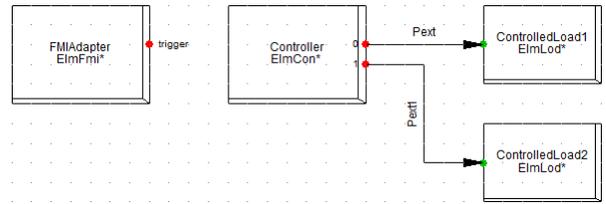

Fig. 1. Example of a composite frame including DSL model FMIAdapter.

To create an FMU from a PowerFactory model, a dedicated Python script has to be executed. The user has to provide all relevant information as command line parameters to this script, such as the FMU model identifier (i.e., the FMU's name), the PowerFactory model, or lists of input and output variable names. For inputs intended to be sent as events via DSL block *FMIAdapter*, a dedicated naming convention has been devised specifying the receiving block and input parameter names.

### B. Generic FMI-based coupling with MOSAIK

The MOSAIK co-simulation framework has been designed for simulation-based testing of novel energy system solutions. Its modular architecture and abstract interface allows the integration of simulation tools from various disciplines, including power systems, information and communication, control, and even economic and social phenomena. This enables users to study dynamic system interactions in a holistic view on their test setup. More information about the architecture and capabilities of MOSAIK are provided in [8].

A crucial concept in the coupling of continuous-time simulation models is the correct handling of the bidirectional data exchange. This aspect has been in the focus of research in the co-simulation of various systems [3], [8]–[10]. For highly dynamic systems, such a coupling is typically conducted iteratively, including several prediction and correction cycles, to arrive at a stable state of the system. In the presented application case, however, such stability issues need no consideration. Furthermore, PowerFactory so far does not support rollbacks of time in the simulation process. Therefore, the data exchange scheduling performed by MOSAIK does not support iterative exchange cycles, but still has to guarantee the avoidance of deadlocks in bidirectional data exchange. MOSAIK provides two options for the handling of cyclic data dependencies: *(i)* either a prioritization among the simulators is established and the data exchange is conducted in a *serial* fashion, or *(ii)* the simulators are executed in *parallel* and provide data for the next time step of each other.

The integration of simulation tools into the MOSAIK environment is typically done via its *Component-API*, which provides a rudimentary structure for the data and methods needed for model interfacing. It has been shown in [9] that a mapping can be established between this Component-API and the more complex FMI standard. An appropriate interface has been established using the FMI++ library so that arbitrary FMUs can be coupled with MOSAIK, which acts as the master algorithm. Since the solver algorithms provided by FMI++ are also employed in the interfaces, it can be used to integrated of both parts of the standard, *FMI for Co-Simulation* and *FMI for Model Exchange*.

All in all, it has been shown that MOSAIK can be employed as a co-simulation master for the given application case, sufficing the requirements of FMI integration and the avoidance of deadlock in the data exchange between FMUs. An additional benefit provided by MOSAIK lies in the flexible creation of co-simulation setups. The so-called Scenario-API allows the script-based definition of executable MOSAIK scenarios based on Python. This makes it easy to conduct parameter studies in the established co-simulation setup, change the number and the coupling of employed simulators, or conduct similar tasks associated with system scaling.

## III. HOLISTIC TESTING AND SCALABLE CO-SIMULATION

ERIGrid devised a systematic approach to define a smart grid test case in such a way that [2], [11]: *(i)* the heterogeneity of present and future smart grids is taken into account, *(ii)* it boasts the implicit inclusion of the Smart Grid Architecture Model (SGAM) and the use case standard IEC 62559-2, and *(iii)* it contains system descriptions at various abstraction levels, amongst others. The latter enables the separation of the *specification* of the test from the *implementation* and *execution* of the test, which allows experiments on various levels of detail and complexity, and even distributing the test over multiple laboratories. This particularly holds for power hardware-in-the-loop (P-HIL) and (co-)simulation, which can for instance be EMT, RMS, or power flow-based. The overarching *scenario*, *use case*, *function under test*, *test criteria*, and *systems under test* can then the same but the experiment design and underlaying assumptions can be different. This makes the experiment very well scalable in terms of system size (i.e., higher granularity of components, scale-out) and simulation size (i.e., no of simulators and models involved, scale up). Results from such (possibly multi-laboratory) experiments can then be used as feedback mechanism for the initial boundary conditions and assumption made [11]. This yields the holistic test case design as shown in Fig. 2.

The co-simulation-based experiment setup focuses on the interaction between the WPP and the main grid, which is the IEEE 9-bus system. The generator at bus 3 has been replaced by the WPP as seen in Fig. 3. The aggregated WPP is connected to the main grid via the point of common coupling (PCC) at bus 3. The function under test is the fault ride through (FRT) capability of the WPP. This function is tested by simulating a 3-phase fault near bus 6. During and after the fault, the WTG must remain connected. Simultaneously, the converter should comply with the FRT curve at the PCC. That is, whenever the voltage-time profile stays above a predefined envelope, the WPP must remain connected [12].

The WTG is assumed to consist of mainly 3 parts: *(i)* the converter controller, *(ii)* the FRT controller, and *(iii)* the machine to grid interface. The converter controller is a vector controller designed as per [13]. The FRT controller acts on top of the converter controller as a finite state machine. The machine-to-grid interface takes the $d$ and $q$-axis projections of the converter current, rotates and scales them, and injects it into the IEEE 9-bus system accordingly, all using the static generator model of PowerFactory. The converter and FRT controllers are designed in MATLAB/Simulink and exported to an FMU using the FMI for model exchange 1.0 specification [4]. The entire co-simulation is orchestrated using the FMI++ toolbox. Two co-simulation cases are developed from this experiment setup.

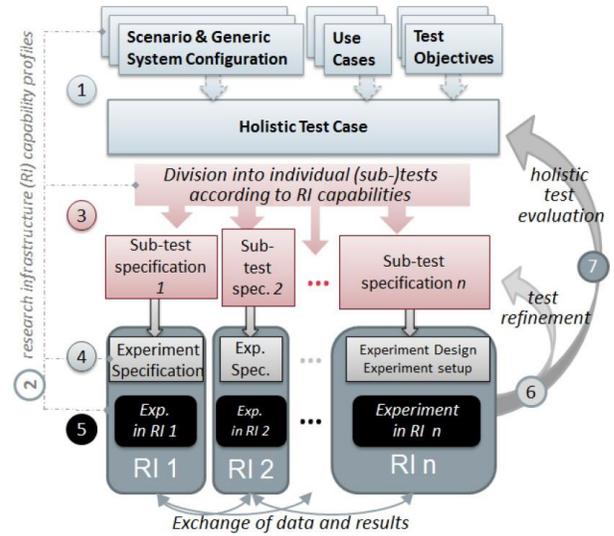

Fig. 2. The ERIGrid holistic test case design steps [2], [11].

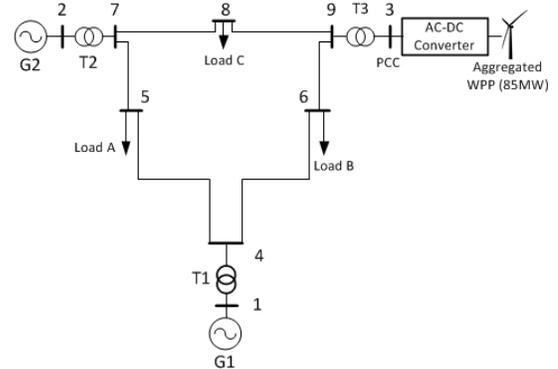

Fig. 3. Modified IEEE 9-bus system with aggregated WPP.

### A. Small scale co-simulation

The starting base for representing the WPP in an aggregated manner is by using a single WTG as shown in Fig. 3. The converter controller consists of two PI controllers and a current limiter. The $q$-axis controller regulates the voltage magnitude of the PCC, whereas the $d$-axis controller maintains the active power reference. The FRT controller alters the control, rate limiter, and current limiter parameters according the voltage dip depth. During the fault, it increases the reactive current output and blocks active power infeed. After fault clearance the active current reference of the WTG is brought back to its prefault point through a maximum ramping rate.

The converter and FRT controllers are exported as FMUs, whereas the PowerFactory model is exported as discussed in Section II-A. Hence, three FMUs are created and the co-simulation is orchestrated through Python/FMI++. The corresponding experiment setup can be seen in Fig. 5. The detailed explanation of the models and setup can be found in [5].

### B. Large scale co-Simulation

Next, a more complex co-simulation setup is conducted in order to analyze the efficiency of the various interfaces, couplings, and controllers developed. In order to significantly increase the complexity, the aggregated WPP is divided into 32 WTGs in an 8x4 array as seen in Fig. 4.

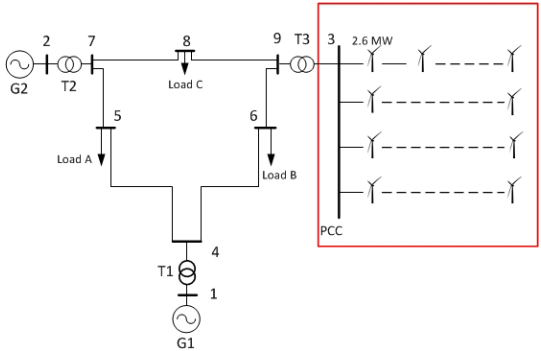

Fig. 4. Modified IEEE 9-bus system with aggregated WPP.

The cable length between the WTG is assumed 700 m, using the cable data of [14] and the network equivalencing approach of [15]. Each WTG is rated 2.6MW; hence the combined power output remains 85MW. Like the previous experiment, each WTG has a set of two controllers (converter and FRT), which are responsible for maintaining the voltage levels at their terminals. Since the voltage control is commonly not centralised the park-level control is assumed to maintain the reactive power at a predefined level. Hence, each WTG applies Q-control during normal operation and voltage magnitude control during FRT. The controllers for each WTG are converted to FMUs along with the PowerFactory model. So, a total of 65 FMUs are created, as shown in Fig. 5. The detailed explanation of the setup and FMU interaction can be found in [16].

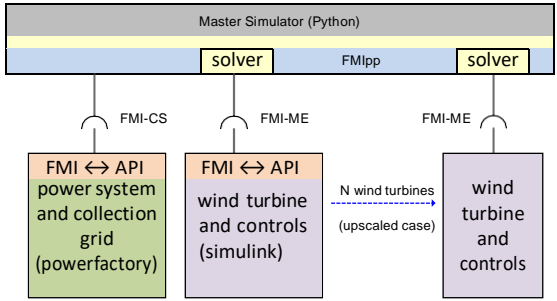

Fig. 5. Co-simulation experiment setup.

## IV. REALIZED SIMULATION STUDIES AND DISCUSSION

### A. Validity of the co-simulation

To test the validity of the co-simulation, the first case study encompasses the comparison between a monolithic PowerFactory simulation and a small scale co-simulation as described in Section III-A. To achieve this, the WTG and FRT models are replicated using PowerFactory standard modeling blocks and the dynamic simulation language. The interactions between the WPP and the main grid mainly manifests themselves at the PCC, hence we will consider the voltage magnitude at the PCC and the output power of the aggregated WPP as the main variables to test the validity of the co-simulation. At $t = 1$s. we simulate a three-phase-to-ground short circuit at bus 6, which causes a voltage dip at the PCC, and is cleared after 180 ms. The time-domain results are shown in Fig. 6 and Fig. 7.

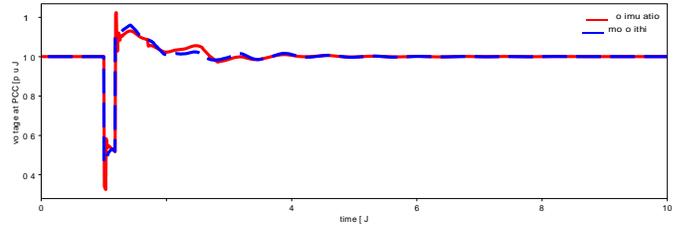

Fig. 6. Voltage at PCC (monolithic vs small scale co-simulation).

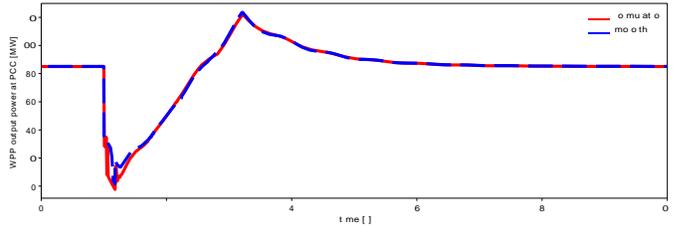

Fig. 7. Output power WPP (monolithic vs small scale co-simulation).

It can be seen that overall simulations trace very similar, especially for the active power exchange between the WPP and the remainder of the grid, which is also the interface location between the subsystems in the co-simulation. The small discrepancy in the voltage magnitude during fault ignition and clearance can be attributed to the relatively simple serial interaction protocol applied in the master simulator.

### B. Scalability of the co-simulation approach

Next, how well the co-simulation approach scales with the models in terms of accuracy, implementation issues, and experienced system phenomena is assessed. For that, the small scale co-simulation with the large scale co-simulation as described in Sections III-A and III-B is compared. As the upscaled co-simulation considers each individual WTG as a set of two FMUs, the grid model from Fig. 3 needed to be extended with the cable array, yielding a slightly different operating point as seen from the PCC. Moreover, the power recovery rate limit has been removed to relieve the computational burden of the simulated case.

The results of a 2 s simulation run are shown in Figs. 8 and 9. During FRT operation the voltage dip depth and power exchange at the PCC differ slightly. This is mainly due to the fact that each WTG experiences a different voltage at its terminals and hence injects a slightly dissimilar reactive current into the collection grid, which is reflected at the PCC. The frequency the oscillations for both cases are similar whereas the damping of the large scale co-simulation is visibly higher. This is a consequence of the additional resistance in the power system as compared to the small scale co-simulation.

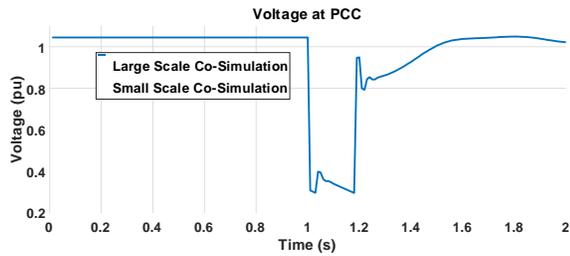

Fig. 8. Voltage at PCC (small scale vs large scale co-simulation).

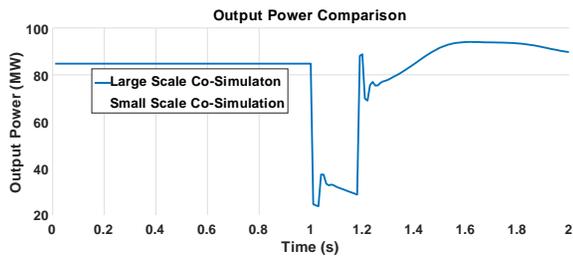

Fig. 9. Output power of WPP (small scale vs large scale co-simulation).

The implementation in the FMI++ package and the master Python script is straightforward. However, as the amount of state variables scales linearly with the amount of models, the execution time does, too. Also, each FMU is associated with a separate numerical integration routine, which brings about computational overhead in the master algorithm, also scaling with system size. This is also reflected into the execution times; the monolithic simulation, small-scale co-simulation, and scaled-up co-simulation took 6, 81, and 105 s respectively. Notwithstanding these small issues, which are common to scaling, the co-simulation approach can be considered well scalable in terms of accuracy and computation effort.

## V. CONCLUSIONS

This work deals with the holistic assessment of grid integration aspects of wind power plants by co-simulation based on the FMI. PowerFactory was lacking such an FMI-based interface for co-simulation and this paper presented an FMI exporter for PowerFactory based on the FMI++ package. The FMI++ and MOSAIK workflow has been presented and applied to a holistic test case in which the fault ride-through of an onshore wind park was studied. First inside a monolithic simulation as a reference, then as a rudimentary co-simulation with three components, and finally as a large-scale co-simulation with a total of 65 functional mock-up units.

The accuracy of the simulations are plausible and visible discrepancies can be attributed to modelling assumptions rather than typical co-simulation factors. This also holds for the scale-up of the study, the execution time of which was only slightly longer than the small-scale experiment (105 s vs 81 s). It is supposed that this is due to the network licensing system of PowerFactory, which was called each synchronisation instance. This is currently part of follow-up studies of FMI++ and MOSAIK integration. The current set of models, simulations, and scripts can be found at [17].

ACKNOWLEDGMENT

This work is supported by the European Community's Horizon 2020 Program (H2020/2014-2020) under project "ERIGrid" (Grant Agreement No. 654113). Further information is available at the corresponding project website erigrid.eu.